\documentclass[a4paper,10pt]{article}
\pdfoutput=1
\usepackage{jheppub}
\usepackage[T1]{fontenc} 
\usepackage{float,tikz-cd,braket,datetime2,graphicx,color,amsfonts,amsthm,bm,bbm,mathrsfs,amsmath,float,pgf,tikz,mathrsfs,xcolor}
\usetikzlibrary{cd}
\definecolor{rindou1}{rgb}{0.4431,0.2862,0.7960}
\definecolor{rindou2}{rgb}{0.0078,0.1215,0.4392}
\definecolor{lapis}{rgb}{0.0.0470,0.2941,0.5568}
\definecolor{burgundy}{rgb}{0.5, 0.0, 0.13}
\usepackage[T1]{fontenc}
\newcommand{\mref}[1]{(\ref{#1})}
\newcommand{\rom}[1]{\mathrm{#1}}
\newcommand{\sq}[1]{`#1'}
\usepackage[english]{babel}
\usepackage[utf8]{inputenc}
\usetikzlibrary{arrows}

\newcommand{\z}{\overline{z}}

\newcommand{\ab}[1]{\langle{#1\rangle}}

\title{Holographic Representations of Supertranslation Eigenstates}

\author{Nikhil Kalyanapuram}

\affiliation{Department of Physics and Institute for Gravitation and the Cosmos, The Pennsylvania State University, University Park PA 16802, USA}
\emailAdd{nkalyanapuram@psu.edu}
\dedicated{Dedicated to the memory of Prof. Steven Weinberg.}

\abstract{We construct by direct computation holographic presentations of supertranslation vacua given a set of supermomenta. To do this, we make use of two-dimensional dual models recently discovered by the author, which encode soft dynamics of gravity at leading order. In particular, the two-dimensional models are used to define soft currents, of which the supertranslation vacua are eigenstates. Operationally, the eigenstates are determined by the bare vacuum state dressed by exponential operators that generalize the dressing due to Faddeev and Kulish.}

\begin{document} 
\maketitle
\flushbottom

\section{Introduction}\label{sec:1}
The rich structure of null infinity in asymptotically flat spacetimes was first pointed out in a series of wonderful papers \cite{Penrose:1962ij,Bondi:1962px,Penrose:1965am,Newman:1968uj,PhysRev.128.2851} by Penrose et al. in the 1960s. Specifically, using the conformal completion of Minkowski space, we can now ask (and sometimes answer) intelligent questions about how radiative data of massless fields are stored and encoded at null infinity. Among the most interesting aspects of the boundaries of asymptotically flat spacetimes is the fact that the diffeomorphism group does not reduce to the Poincar\'e group as one might expect from the condition of asymptotic flateness. Rather, the asymptotic symmetry group is an infinite-dimensional extension of the Poincar\'e group named after Bondi, Metzner and Sachs (BMS). The BMS group, put most simply, consists of the conventional Lorentz group and translations on null infinity that are angle dependent. Parameterized by an infinite set of integers when expanded in spherical harmonics, such translations form an infinite-dimensional generalization of the diffeomorphism group of flat spacetime.

While much of this work was done in largely classical contexts, it is clear even at the outset that the study of such asymptotic symmetries must be important in the larger context of quantum field theory as well, specifically in the study of quantum gravity. The reason for this is simply the fact that in the absence of a truly microscopic theory of quantum gravity, it is already interesting to study the behaviour of quantum fields in the presence of gravity. The simplest such case of physical interest would be that of asymptotically flat spacetimes. In this light, understanding the quantization of quantum fields at null infinity poses a very interesting question. The quantization of massless fields at null infinity was carried out in depth by Ashtekar et al. \cite{Ashtekar:1979xeo,Ashtekar:1981bq,Ashtekar:1981hw,Ashtekar:1981sf,Ashtekar:2018lor}, who found that long-range radiative data was stored in an intricate fashion at null infinity. Specifically, degrees of freedom that corresponded to arbitrarily long wavelengths (or arbitrarily small energies) are collected at various points on null infinity, acting as signals of the dynamics in the bulk that are otherwise inaccessible. 

The study of quantum field theory deep in the infrared has a long history, and has been studied by a number of authors over the years \cite{PhysRev.52.54,PhysRev.76.790,Yennie:1961ad,Kinoshita:1962ur,Lee:1964is,Kibble:1968sfb,Kibble:1969ip,Kibble:1969ep,Kibble:1969kd,Weinberg:1964ew,Weinberg:1965nx,DeWitt:1967uc,Jauch:1976ava}. Put most concisely, scattering amplitudes in quantum field theories with massless quanta are known to be divergent deep in the infrared, due to uncontrolled radiation of massless particles of arbitrarily low energy. It is known however that when certain assumptions are made (such as the eikonal condition), these infrared divergences can be eliminated at the level of the differential cross section, or if one is careful, even at the level of the amplitudes themselves \cite{Chung:1965zza,Kulish:1970ut,Hannesdottir:2019rqq,Hannesdottir:2019opa}. That being said, its natural to ask if there is any connection between the infrared aspects of quantum field theory and the subtleties of quantizing massless theories at null infinity. 

It turns out that this is indeed the case. It has become clear due to recent analyses\footnote{The literature on this subject is unwieldy---the reader is pointed towards \cite{Strominger:2017zoo,Kalyanapuram:2021tnl} and the references therein.}
that the soft factorization of scattering amplitudes in conventional quantum field theory are closely related to asymptotic symmetries at null infinity. For gravity in particular, it is now known that the generators of supertranslations give rise to an infinite set of conservation laws. For any choice of supertranslation, it is required that the generator of the symmetry transformation commute with the $S$-matrix. It turns out that a specific class of such identities is equivalent to the statement that the $S$-matrx factorizes upon radiation of a single soft graviton. This relation establishes that the soft theorem in conventional quantum field theory is not just a natural consequence of unitarity and invariance under gauge transformations as one might expect. Rather, it comes out of the fact that there is a natural symmetry structure of spacetime itself in the presence of gravity, of which the soft theorem is simply one presentation.

With this in mind, we can now ask if it is possible to reformulate quantum field theory on asymptotically flat spacetimes in accordance with the approach taken by Weinberg in his series of seminal papers of Feynman rules \cite{Weinberg:1964cn,Weinberg:1964ev}. In those papers, Weinberg constructs one-particle states by having recourse directly to their presentations as irreducible representations of the Lorentz group. Indeed, by ensuring that the states obey the right transformation rules under the action of the translational and connected parts of the full Poncair\'e groups, the resulting irreducible representations turn up as precisely the ones cooresponding to single-particle states in more conventional quantum field theory, from which a unified set of Feynman rules could be inferred. Indeed, performing this analysis by taking as our global group of symmetries the BMS group instead of the Poincar\'e group would represent a significant step towards realising a macroscopic theory of quantum gravity.

It turns out that this question was largely answered in a series of papers by McCarthy in the 1960s \cite{McCarthy1,McCarthy2,McCarthy3,McCarthy4}. Somewhat old-fashioned in their approach, the finding of these papers of most interest to us is the fact that the character group corresponding to BMS symmetries essentially correspond to conventional momenta associated to four-dimensional translations along with an infinite set of numbers that can be thought of as `supermomenta', suggestively indicating that they are dual to supertranslations along the celestial sphere. In modern language, the analogues of conventional single particle states when the global symmetry group is the BMS group are states labelled as usual by four momenta and spin, along with an infinite set of numbers corresponding to so-called supertranslation charges. Our task in the present work is to tie this larger question with the more modern puzzle of flat-space holography.

In the simplest of terms, flat-space holography refers to the generic idea that a true quantum theory of gravity will not make reference to the four-dimensional bulk that is the traditional area of gravitational interactions. Rather, the fundamental idea behind flat-space holography is the claim that all the dynamical data characterizing interactions and scattering in the four-dimensional bulk will be \emph{equivalently} described by a lower-dimensional theory living on one of the boundaries of the larger four-dimensional spacetime. In the context of those spacetimes that are asymptotically flat, the essential hope is that a corresponding theory of quantum gravity would be a lower dimensional theory living naturally on null infinity or one of the boundaries thereof.

While this problem remains more or less intractable for the full dynamics of four-dimensional gravity, it turns out to be considerably simpler to ask intelligent questions as long as we restrict our attention to the soft sector of gravity. Indeed, in a recent series of papers due to the present author \cite{Kalyanapuram:2020epb,Kalyanapuram:2021bvf,Kalyanapuram:2021tnl}, it was shown that in the case of QED and gravity, the leading order soft sectors are dynamically captured by theories that are entirely two dimensional. In the case of QED, this theory turns out to be the simple case of a free boson on the sphere, while in the case of gravity this theory is that of a two-dimensional field obeying a biharmonic equation. In particular, since it is known that there are no divergences in gravity at leading order \cite{Weinberg:1965nx,DeWitt:1967uc,Naculich:2011ry,Akhoury:2011kq,Beneke:2012xa}, the two-dimensional dual for the soft sector along with a putative theory\footnote{There is some indication that a full theory for the hard part would likely be three dimensional. See for example \cite{Kalyanapuram:2020aya} for considerations at tree level.} for the hard part would constitute a macroscopic model of quantum gravity on asymptotically flat space.

In accordance with the broad goal of finding a manifestly holographic description of gravity, we consider the question of irreducible representations of the BMS group in a more modern fashion. Specifically, we will study this problem in the context of the two-dimensional duals that we now know describe the soft part of gravity at leading order. Indeed, since we now know that the soft theorems are really invariance under supertranslations in disguise, we will recast the entire problem of constructing supertranslation eigenstates in terms of the two-dimensional models of \cite{Kalyanapuram:2020epb,Kalyanapuram:2021bvf,Kalyanapuram:2021tnl}. In concrete terms, we will give a prescription of how to construct eigenstates of supertranslations given a list of supermomenta in terms of the dual two-dimensional model of soft gravity.

In the interest of making this paper somewhat self-contained, we have provided in Section \ref{sec:2} a review of the relationship between supertranslation Ward identities and the leading order soft graviton theorem. The main body of this paper is contained in Section \ref{sec:3}, where after a brief review of the two-dimensional dual model for soft gravitons, we construct by direct computation eigenstates for supertranslations. While we first consider the simplest such construction, a refined version is studied after introducing an analogue of chiral splitting for the two-dimensional dual model. We end the paper with a discussion of further issues in Section \ref{sec:4}.

Throughout this paper, we make a few assumptions regarding the choice of bases and mass-shell condition for the momenta of hard particles participating in the scattering process. We will work in the momentum basis, while we take all particles in the scattering process to be massless, but of finite energy, in accordance with which the momentum $p_{i}$ of an external state will be expanded as

\begin{equation}
    p_{i} = \omega_{i}(1+z_{i}\z_{i},z_{i}+\z_{i},-i(z_{i}-\z_{i}),1-z_{i}\z_{i}).
\end{equation}
Here, the energy of the particle has been denoted by $\omega_{i}$ while the remaining two components of the momentum vector have been mapped onto $\mathbb{CP}^{1}$, so that $z_{i}$ and $\z_{i}$ are punctures on the celestial sphere. Generically, we do not require that the punctures $z_{i}$ and $\z_{i}$ be complex conjugate to each other. 

To any massless particle we may associate a pair of polarization vectors, which are defined according to

\begin{equation}
    \epsilon^{+}_{i} = \frac{1}{\omega_{i}}\partial_{z_{i}}p_{i}
\end{equation}
with the negative helicity counterpart obtained by replacing $z_{i}$ with $\z_{i}$. We use this convention when we rewrite soft theorems in celestial coordinates. 

\section{Ward Identities and Supertranslation Eigenstates}\label{sec:3}
The simplest way to understand supertranslation eigenstates is to go back to the presentation of the leading soft graviton theorem in terms of the supertranslation Ward identity. In particular, the decomposition of the Noether charge makes very explicit the definition of the supertranslation eigenstates. In this section, we will briefly review this procedure and follow with the main results of this paper in the next section.

Recall that the leading soft graviton theorem is equivalent to the statement that the generator of supertranslations commutes with the $S$-matrix evaluated between bare states. By bare states, we do \emph{not} refer to states unregulated in the UV, but those which have not been dressed using the method of Faddeev and Kulish. More concretely, the leading soft graviton theorem is equivalent to saying that

\begin{equation}
    \braket{\rom{out}|[Q_{f},S]|\rom{in}} = 0
\end{equation}
for any function $f(z,\z)$, defined on the celestial sphere. This function is the so-called \emph{dressing function}, which smears the charge $Q(z,\z)$ defined independently at all points on $\mathbb{CP}^{1}$ over the celestial sphere. 

The leading soft graviton theorem (for either helicity choice) is obtained by making particular choices for the function $f$. The correct choice can be gleaned by making note of the explicit analytical expansion of charge, which takes the form

\begin{equation}
    Q(z,\z) = Q^{\rom{soft}}(z,\z) + Q^{\rom{hard}}(z,\z)
\end{equation}
where

\begin{equation}
   Q^{\rom{soft}}(z,\z) = \int_{-\infty}^{\infty}\left(D_{\z}^{2}a_{+}(u,z,\z)+\rom{c.c} \right)du,
\end{equation}
and

\begin{equation}
    Q^{\rom{hard}}(z,\z) = \int_{-\infty}^{\infty}T_{uu}(z,\z)du.
\end{equation}
For those readers unfamiliar with the notation, $u$ is the retarded coordinate at future null infinity. The \sq{soft} and \sq{hard} labels are meant to be suggestive; the soft part of the supertranslation charge---owing to the retarded time integral---creates a soft photon state at the point $(z,\z)$, smeared using the function $D^{2}_{\z}f(z,\z)$ (or its conjugate). The hard function on the other hand is more easily evaluated; we have

\begin{equation}\label{eq:2.5}
    \braket{\rom{out}|[Q^{\rom{hard}}_{f},S]|\rom{in}} = \sum_{i=1}^{n}\eta_{i}\omega_{i}f(z_{i},\z_{i})\braket{\rom{out}|S|\rom{in}}
\end{equation}
where as usual $\eta_{i}=\pm 1$ for incoming and outgoing states respectively. The structure of this action immediately indicates to us what our choice for the function $f(z,\z)$ has to be. Consider the case of the soft graviton theorem when there is a positive helicity graviton being radiated. We know that in stereographic coordinates, when the soft momentum $q$ is expanded as

\begin{equation}
    q = \lim_{\omega\rightarrow 0}\omega (1+z'\z',z'+\z',-i(z'-\z'),1-z'\z')
\end{equation}
the soft factor becomes

\begin{equation}
    \mathcal{S}^{0}(z',\z') = \sum_{i=1}^{n}\kappa\eta_{i}\omega_{i}\frac{\z'-\z_{i}}{z'-z_{i}}.
\end{equation}
Clearly, the following choice for the function $f$

\begin{equation}
    f(z,\z) = \frac{\z'-\z}{z'-z}
\end{equation}
turns equation \mref{eq:2.5} into the positive helicity leading soft graviton theorem. Furthermore, the soft part of the charge becomes especially simple with this choice of the function $f$; we obtain

\begin{equation}
    Q_{f}^{\rom{soft}} = \lim_{\omega\rightarrow 0}\omega a_{+}(\omega,z',\z')
\end{equation}
where the operator $a_{+}(\omega,z',\z')$ creates an outgoing positive helicity soft graviton. The fact that the supertranslation charge commutes with the $S$-matrix is now readily seen to be equivalent to the soft theorem; the soft part of the commutator is the $S$-matrix element in the presence of one soft graviton, while the hard part of the commutator is Weinberg's well-known result.

The interpretation of the soft theorem in terms of supertranslation symmetries has a broader implication. Most simply, it indicates that the vacuum in a theory with gravitation is really quite degenerate, with superselection sectors identified by their so-called \sq{soft charge}. What we mean by this is the following - note that the fact that the supertranslation charge commutes with the $S$-matrix is not equivalent to stating that it annihilates the vaccum. Indeed, most of the time we have

\begin{equation}
    Q_{f}\ket{0} \neq 0.
\end{equation}
At the same time, we can recall that the hard part of the charge naturally annihilates the vacuum (assuming of course that the vacuum has zero energy\footnote{The extent to which this assumption can be made in the context of a gravitational theory is debatable. For the purposes of the present work we stick to this point of view.}). The immediate result implied by this is that it is actually the soft part of the charge that is responsible for acting non-trivially on the vacuum. Under the assumption that the vacuum is an eigenstate of the soft part of the generator of supertranslation, the corresponding eigenvalue is the soft charge of that state. 

It's possible to make all of this more precise if we observe that the soft charge has been smeared over by the function $f(z,\z)$, which we have---in terms of stereographic coordinates---taken to be a function on the celestial sphere. Making a coordinate transformation to conventional sphere variables $(\theta,\phi)$ makes it possible to expand the function in a well-known basis, namely that of the spherical harmonics, according to

\begin{equation}
    f(\theta,\phi) = \sum_{\ell=0}^{\infty}\sum_{m=-\ell}^{\ell}f_{\ell m}Y_{\ell m}(\theta,\phi).
\end{equation}
As a consequence of this, the soft charge smeared using $f$ can itself be expanded in the spherical basis, with the charge now being controlled by $Q_{\ell m}^{\rom{soft}}$, obtained by smearing $Q(z,\z)$ with $Y_{\ell m}(\theta,\phi)$. 

With this basis in hand, it becomes clear that \emph{any} soft charge can be expanded in terms of the basis operators $Q_{\ell m}^{\rom{soft}}$, which tells us that any supertranslation eigenstate is uniquely determined by the (countably) infinite set of conditions given by

\begin{equation}
    Q_{\ell m}^{\rom{soft}}\ket{\lbrace{s_{\ell m}\rbrace}} = s_{\ell m}\ket{\lbrace{s_{\ell m}\rbrace}}.
\end{equation}
Importantly, we have to keep in mind that none of these states contain particles with finite energy. Each of these states is nominally of zero energy; together they span all superselection sectors of the vacuum as a whole, which accordingly ends up being highly degenerate. 

The question at hand now is the following---is it possible to construct more explicitly, preferably in terms of specific field configurations, these eigenstates of the soft charge? To answer this question, it turns out that there is a particularly clean method of constructing these eigenstates using a two-dimensional dual model for soft theorems developed by the author, in terms of a two-dimensional field that obeys a biharmonic equation. In the following section, we briefly recall this model and proceed to explain how it makes possible a clean definition of a soft charge eigenstate.

\section{Soft Eigenstates and Two-Dimensional Dual Fields}\label{sec:2}
Let us start with a brief overview of the two-dimensional model due to the author as given in \cite{Kalyanapuram:2020epb} which provides a dual description of the soft sector of gravity at leading order. The model is motivated by noticing that the transformation given by

\begin{equation}
    f(z,\z) \longrightarrow f(z,\z) + a_{1} + a_{2}z + a_{3}\z + a_{4}z\z = f(z,\z) + \delta f(z,\z)
\end{equation}
has a trivial effect on the soft charge due to the fact that

\begin{equation}
    D_{z}^{2}\delta f(z,\z) = D_{\z}^{2}\delta f(z,\z)=0
\end{equation}
for constant $a_{i}$ (the commutation of $Q_{\delta f}$ with the $S$-matrix simple becomes the global conservation of four-momentum). Its natural to expect that a two-dimensional dual for soft gravitons would enjoy the same symmetries on the celestial sphere. Indeed, the natural choice for the corresponding action is simply

\begin{equation}
    I = \int dz\wedge d\z \left(D_{z}^{2}\sigma(z,\z)D_{\z}^{2}\sigma(z,\z)\right)
\end{equation}
where $\sigma(z,\z)$ is a field on the celestial sphere, which as we shall see encodes the dynamics of the leading order soft theorem. The first step to noticing this is to recognize that we have the following operator product expansion

\begin{equation}
    \sigma(z,\z)\sigma(z',\z') \sim |z-z'|^{2}\ln|z-z'|^{2}
\end{equation}
due to the fact that the right side of this equation is precisely the Green's function of the square of the Laplacian on the celestial sphere. Additionally, due to the global symmetries of this model, we have two Noether currents, given by

\begin{equation}
    j(z,\z) = D_{z}^{2}\sigma(z,\z)
\end{equation}
and its counterpart, obtained by making the replacement $D_{z}\rightarrow D_{\z}^{2}$. 

These currents, as it turns out, are rather important. Although they are just the Noether currents corresponding to invariance under translations on the celestial sphere, they encode soft degrees of freedom. Indeed, seeing this is just a matter of observing that we have the operator product expansions---

\begin{equation}
    D_{z}^{2}\sigma(z,\z)\sigma(z',\z') \sim \frac{\z-\z'}{z-z'},
\end{equation}
and

\begin{equation}
    D_{\z}^{2}\sigma(z,\z)\sigma(z',\z') \sim \frac{z-z'}{\z-\z'},
\end{equation}
which are identified as the soft factors for positive and negative helicity graviton radiation respectively. The next step is to set up scattering amplitudes in a fashion that ensures that one insertion of the soft current into the matrix element results in multiplication by the above factors. The method to do this is as follows. Suppose external states are created by asymptotic operators labelled by energy and celestial coordinates. We denote such operators by $\mathcal{O}(\omega,z,\z)$. We are now looking for a modification of such operators, denoted by $\widetilde{O}$ such that we have

\begin{equation}
    t(z,\z)\widetilde{\mathcal{O}}(\omega_{i},z_{i},\z_{i}) \sim \omega_{i}\frac{\z-\z_{i}}{z-z_{i}}\widetilde{\mathcal{O}}(\omega_{i},z_{i},\z_{i}). 
\end{equation}
The right choice of the modification is informed by the fact that we have the following identity---

\begin{equation}
    t(z,\z)\exp(i\kappa\eta_{i}\omega_{i}\sigma(z_{i},\z_{i})) \sim \kappa\eta_{i}\omega_{i}\frac{i}{\pi}\frac{\z-\z_{i}}{z-z_{i}}\exp(i\kappa\eta_{i}\omega_{i}\sigma(z_{i},\z_{i})).
\end{equation}
In accordance with this fact, we define

\begin{equation}
    \widetilde{\mathcal{O}}(\omega_{i},z_{i},\z_{i}) = \exp(i\kappa\eta_{i}\omega_{i}\sigma(z_{i},\z_{i}))\mathcal{O}(\omega_{i},z_{i},\z_{i}).
\end{equation}

The lesson of this review is simply that the two-dimensional theory governing the evolution of $\sigma$ provides a holographic prescription for constructing soft dressings for hard external states, entirely analogous to the ones constructed by Faddeev and Kulish. The question now is---how do we construct eigenstates of the soft current, namely those of $j$ (or its conjugate current)? For this purpose, let us construct smeared versions of the currents and dressings according to the definitions

\begin{equation}
    t_{f} = \int f(z,\z)D_{z}^{2}\sigma(z,\z) dz\wedge d\z
\end{equation}
and

\begin{equation}
    \mathcal{W}_{g}(\omega) = \exp\left(\int g(z,\z)\sigma(z,\z)dz\wedge d\z\right).
\end{equation}
Given these, let us first observe that we have the following operator product expansion

\begin{equation}
    t_{f}\mathcal{W}_{g}(\omega) \sim \Lambda_{\omega}(f,g)\mathcal{W}_{g}(\omega)
\end{equation}
for

\begin{equation}\label{eq:3.14}
    \Lambda_{\omega}(f,g) \sim \int d^{2}zd^{2}z' \left(f(z,\z)D_{z}^{2}G(z,\z;z',\z')g(z',\z')\right)
\end{equation}
where we have defined for brevity

\begin{equation}
    G(z,\z;z',\z') = |z-z'|^{2}\ln|z-z'|^{2}.
\end{equation}
Answering the question posed earlier is now a matter of choosing the right basis. First, the form of equation \mref{eq:3.14} is put in more covariant form by noting that we have

\begin{equation}\label{eq:3.16}
    \Lambda_{\omega}(f,g) \sim \int d\Omega d\Omega' \left(f(\Omega)n^{A}n^{B}D_{A}D_{B}G(\Omega;\Omega')g(\Omega')\right)
\end{equation}
where $\Omega$ denotes a coordinate system on the sphere, while $n^{A}$ is a null vector, assuming the following decomposition of the sphere metric

\begin{equation}
    \gamma^{AB} = n^{(A}\overline{n}^{B)}.
\end{equation}
For the special choice of

\begin{equation}
    f(\Omega) = Y_{\ell m}(\Omega)
\end{equation}
and the choice

\begin{equation}\label{eq:3.19}
    g(\Omega') = \sum_{\ell m}s_{\ell m}\overline{n}^{A}\overline{n}^{B}D_{A}D_{B}Y^{*}_{\ell m}(\Omega')
\end{equation}
its easy to see that the form of $\Lambda$ just becomes\footnote{We remark that the integration by parts done to achieve this form is only possible because the spin weights of the functions inside the $\Omega'$ integral add up to zero.}

\begin{equation}
    \Lambda_{\omega}(f,g) = \sum_{\ell, m}s_{\ell' m'} \int d\Omega d\Omega' Y^{*}_{\ell m}(\Omega)\delta^{2}(\Omega-\Omega')Y_{\ell' m'}(\Omega')
\end{equation}
which becomes $\sim s_{\ell m}$. 

Pausing for a moment, its helpful to make this discussion a little more precise. We have defined a dressed version of the current for positive helicity soft radiation and called it $j_{f}$. Following this, we showed that for the specific choice of the function \mref{eq:3.19}, the smeared version of the dressing operator, when viewed in terms of a state-operator correspondence, amounts to an eigenstate of the soft charge with eigenvalues $s_{\ell m}$. More concretely, 

\begin{equation}
    \ket{\lbrace{s_{\ell m}\rbrace}_{+}} \Longleftrightarrow  \exp\left(\int g(z,\z)\sigma(z,\z)dz\wedge d\z\right)\ket{0}.
\end{equation}
The $+$ subscript is used to denote the fact that this is an eigenstate of the positive helicity charge. The question of whether or not this can be refined into an eigenstate of the positive and negative helicity charges thus remains open. Although we don't yet have a definitive answer to this, we will now discuss a possible resolution to this problem. To do this, let us consider the simpler example of the free boson in two dimensions, to import its inclusion of chirality into our model.

The free boson is characterized by a single scalar field $\phi$ that obeys the operator product expansion

\begin{equation}
    \phi(z,\z)\phi(z',\z') \sim \ln|z-z'|^{2}.
\end{equation}
The free boson is known to be an achiral theory, which does not discriminate between left- and right-moving components. This is seen by performing the decomposition

\begin{equation}
    \phi(z,\z) = \phi_{+}(z) + \phi_{-}(\z)
\end{equation}
where

\begin{equation}
    \phi_{+}(z)\phi_{+}(z') \sim \ln(z-z')
\end{equation}
and an anti-holomorphic analogue for $\phi_{-}$. Taking our cue from this, let us define the following decomposition of our model according to

\begin{equation}
    \sigma(z,\z) = \sigma_{+}(z,\z) + \sigma_{-}(z,\z)
\end{equation}
where

\begin{equation}
    \sigma_{+}(z,\z)\sigma_{+}(z',\z') \sim |z-z'|^{2}\ln(z-z')
\end{equation}
and

\begin{equation}
    \sigma_{-}(z,\z)\sigma_{-}(z',\z') \sim |z-z'|^{2}\ln(\z-\z').
\end{equation}
This allows us to generalize the currents to incorporate the expected chirality exhibited by the positive and negative helicity theorems; we have for the positive helicity current now

\begin{equation}
    t^{+}_{\ell m} = \int d\Omega Y_{\ell m}(\Omega)n^{A}n^{B}D_{A}D_{B}\sigma_{+}(\Omega),
\end{equation}
and the negative helicity counterpart is obtained by making the replacements $n^{A}\rightarrow \overline{n}^{A}$ and $\sigma_{+}\rightarrow \sigma_{-}$. With these, let us look at the state defined by

\begin{equation}
    \ket{\lbrace{s_{\ell m}\rbrace}} = \exp\left(\ab{g_{-},\sigma_{+}}+\ab{g_{+},\sigma_{-}}\right)
\end{equation}
where the angle brackets just denote the inner product on the sphere, while the functions $g_{\pm}$ have the expansions

\begin{equation}
    g_{+}(\Omega) = \sum_{\ell m}s_{\ell m}\int d\Omega \overline{n}^{A}\overline{n}^{B}D_{A}D_{B}Y^{*}_{\ell m}(\Omega),
\end{equation}
and
\begin{equation}
    g_{-}(\Omega) = \sum_{\ell m}s_{\ell m}\int d\Omega n^{A}n^{B}D_{A}D_{B}Y^{*}_{\ell m}(\Omega).
\end{equation}
For these choices, it is clear that $\ket{\lbrace{s_{\ell m}\rbrace}}$ satisfies

\begin{equation}
    t^{\pm}_{\ell m}\ket{\lbrace{s_{\ell m}\rbrace}} = s_{\ell m}\ket{\lbrace{s_{\ell m}\rbrace}}
\end{equation}
which are consequently the supertranslation eigenstates that we wanted to define. 

Finally, let us briefly comment on how a similar picture can be understood regarding the vacuum in QED, and the degeneracy thereof due to soft excitations. In the case of QED, it is known that a similar asymptotic model that captures the soft degrees of freedom exists, namely that of the free boson---

\begin{equation}
    I^{\rom{QED}} = -\int dz\wedge d\z \left(D_{z}\varepsilon(z,\z)D_{\z}\varepsilon(z,\z)\right) 
\end{equation}
where this time, the field $\varepsilon$ obeys the harmonic equation. The fact that this is the correct theory to describe soft interactions at leading order is gleaned from the operator product expansion

\begin{equation}
    D_{z}\varepsilon(z,\z)\varepsilon(z',\z') \sim \frac{1}{\pi}\frac{1}{z-z'}
\end{equation}
and its antiholomorphic counterpart. This is the singularity structure exhibited by the leading soft photon theorem. Indeed, by defining dressed operators for charged external operators denoted by $\mathcal{O}(e_{i},z_{i},\z_{i})$ according to

\begin{equation}
    \widetilde{\mathcal{O}}(e_{i},z_{i},\z_{i}) = \exp\left(i\eta_{i}e_{i}\varepsilon(z_{i},\z_{i})\right)\mathcal{O}(e_{i},z_{i},\z_{i})
\end{equation}
the action of the current $D_{z}\varepsilon(z,\z)$ on this dressed state simply produces the soft factor due to the radiation of a soft photon of positive helicity by the $i$\textsuperscript{th} external hard state. That being said, defining eigenstates of the soft currents once again requires an appropriate refinement of this dressing operator. By defining ($\varepsilon_{\pm}$ are just the chiral components of the field $\varepsilon$)

\begin{equation}
    \ket{\lbrace{a_{\ell m}\rbrace}} = \exp\left(\ab{h_{-},\varepsilon_{+}}+\ab{h_{+},\varepsilon_{-}}\right)
\end{equation}
where

\begin{equation}
    h_{+}(\Omega) = \sum_{\ell m}s_{\ell m}\int d\Omega \overline{n}^{A}D_{A}Y^{*}_{\ell m}(\Omega),
\end{equation}
and
\begin{equation}
    h_{-}(\Omega) = \sum_{\ell m}s_{\ell m}\int d\Omega n^{A}D_{A}Y^{*}_{\ell m}(\Omega)
\end{equation}
we have

\begin{equation}
    j^{\pm}_{\ell m} \ket{\lbrace{a_{\ell m}\rbrace}} = a_{\ell m}\ket{\lbrace{a_{\ell m}\rbrace}}
\end{equation}
for

\begin{equation}
    j^{+}_{\ell m} = \int d\Omega n^{A}D_{A}\varepsilon_{+}(\Omega)Y_{\ell m}(\Omega)
\end{equation}
with its antiholomorphic counterpart defined accordingly, by carrying out the replacements $n^{A}\rightarrow \overline{n}^{A}$ and $\varepsilon_{+}\rightarrow \varepsilon_{-}$. 

\section{Discussion}\label{sec:4}
In this short paper, our focus has been to provide concrete representations of eigenstates of supertranslation generators on null infinity in terms of two-dimensional fields. We did this by making use of two-dimensional dual models for the leading order soft sector in gravity developed by the author in a series of recent papers. By reconstructing the soft part of the supertranslation charge in terms of the two-dimensional dual of the soft graviton field, it became possible to construct by direct computation the eigenstate of the soft charge. Since the soft charge is defined locally in an angle-dependent fashion, this was done by expanding the soft charge in terms of spherical harmonics, leading to a representation of the supertranslation eigenstates in terms of the corresponding eigenvalues.

The fact that the supertranslation eigenstates have been constructed using an analogue of the conventional dressing operator due to Faddeev and Kulish, specifically considering their exponential form, tells us that it would be interesting to study how such eigenstates are related to the dressed operators denoted by $\widetilde{\mathcal{O}}$. In particular, it might be revealing to study the extent to which the soft dressings of external states are equivalent to simply adding supertranslation degrees of freedom to one-particle states.

Throughout this paper, we have made no attempt to construct states of any immediate physical value. In more mundane language, we haven't specified any constraints on our choice of functions that we use to defined dressed soft charges. While at first glance this may not seem like an important question, we recall that even in the case of normal four-dimensional translations, elementary simplifications (such as the non-vanishing of the three-gluon vertex in QCD) are obtained by generalizing real momenta to the complex plane while remaining on the mass-shell. This question becomes more relevant if we insist that the supertranslation eigenstates be orthogonal for different choices of supermomenta. Indeed, this will certainly be the case under the assumption that the smeared soft charges are Hermitian. Studying this point further will be important when it becomes necessary to construct complete bases of supertranslation eigenstates.

We saw that in the same manner as supertranslation eigenstates are defined for vacua refined according to their supertranslation charges it is possible to refine the traditional vacuum of QED into superselection sectors defined according to their soft charges. In the case of QED however the asymptotic symmetry group is composed of large gauge transformations, corresponding to changes in phase which are angle-dependent. It is known that such large gauge transformations and the associated refinement of the QED vacuum lead to the breaking of Lorentz invariance \cite{Frohlich:1978bf,Frohlich:1979uu,Balachandran:2013wsa,Balachandran:2014hra}. Givent hat our construction has been entirely two-dimensional in the present work, it might be interesting to see whether or not our construction can encode the patterns of Lorentz symmetry-breaking found in the conventional QED analysis.

From the calculations of the previous section, the reader will have noticed that the dynamical two-dimensional dual models for soft QED and gravity are related by two simple transformations, namely the replacements of the derivative operators by their squares---

\begin{equation}
    (\partial,\overline{\partial}) \longrightarrow (\partial^{2},\overline{\partial}^{2})
\end{equation}
and the replacement $e_{i}\longrightarrow \kappa\omega_{i}$. To readers familiar with the proposed duality between color and kinematics, this would look like a celestial avatar of the so-called double copy relation between gauge theory and gravity. However, considering the fact that the dual theory for soft gravity can actually be found by carefully studying the boundary dynamics of conventional general relativity \cite{Nguyen:2021qkt}, it is more likely that this double copy arises out of the more standard double copy between gauge theory and gravity known to exist at tree level. Performing the analysis of \cite{Nguyen:2021qkt} for the case of QED will tell us more about this point. More interestingly however, the double copy has on occasion been put forth as a potential \emph{definition} of gravity. This suggestion largely stems from the hope that observables in gravity can be given a meaning independent of the complex dynamics of the Einstein-Hilbert action by application of the double copy to the more well-understood dynamics of gauge theory.

While a better understanding of the double copy construction would be of theoretical interest, when viewed from the larger perspective of flat space holography it may not serve as an adequate description of gravity in its entirety. The reasons for this are two-fold. First, recent studies have indicated that the double copy may be a low-energy result of the intricate structure of amplitudes in superstring theory \cite{Mizera:2017cqs,Mizera:2019gea,Mizera:2019blq,Kalyanapuram:2021xow,Kalyanapuram:2021vjt}. Accordingly, studying gravity from the point of view of the double copy, at least insofar as we are concerned about low-energy phenomena, might not be the most general approach (for some recent studies of the double copy at null infinity see \cite{Casali:2020vuy,Casali:2020uvr,Kalyanapuram:2020aya}). 

Second, it has been suggested by recent papers (due to a number of authors) \cite{Laddha:2020kvp,Chowdhury:2020hse} that there is something intrinsically holographic about gravitation once certain assumptions are made. When dealing with the asymptotically flat case in particular, when the perturbative result that every operator that effects a transition from one supertranslation eigenstate to another is contained in the algebra of operators at $u=-\infty$ at null infinity is taken to be a feature of the full microscopic theory, it turns out that the past boundary of future null infinity holds information holographically. Concretely, any operator to the future of this boundary can be recovered using operators at $u=-\infty$. This observation is important because---among other things---it seems to indicate that there is no problem in relating information stored in a black hole to that contained in Hawking radiation; the information is always present at the past boundary of future null infinity. It seems unclear to the author how this intrinsically holographic feature of gravity (if it is indeed a property of the microscopic refinement of QFT on asymptotically flat spacetimes) can be obtained from a double copy definition of gravity. Understanding this point further is likely to be important in the context of developing a better description of quantum gravity on asymptotically flat spaces.

\section*{Acknowledgements} 
I thank Jacob Bourjaily and Alok Laddha for their comments. Partial support for this research is due to ERC Starting Grant (No. 757978) and a grant from the Villum Fonden (No. 15369). This work has been supported in part by the US Department of Energy (No. DE-SC00019066).



\bibliographystyle{utphys}
\bibliography{v1.bib}

\end{document}